\documentclass[aps,twocolumn,nofootinbib,showkeys,showpacs,superscriptaddress]{revtex4-1}
\usepackage{amssymb}
\usepackage{amsmath}
\usepackage{amstext}
\usepackage{amsfonts}
\usepackage{bbold}
\usepackage{bm}
\usepackage{graphics}
\usepackage{mathtools}
\usepackage{lmodern}
\usepackage{graphicx}
\usepackage{hyperref}
\usepackage{xcolor}
\usepackage[export]{adjustbox}

\begin{document}

\title{Dark Matter, Dark Photon and Superfluid He-4 from Effective Field Theory} 

\author{Andrea~Caputo}
\affiliation{Instituto de Fisica Corpuscular, Universidad de Valencia and CSIC, Edificio Institutos Investigacion, Catedratico Jose Beltran 2, Paterna, 46980 Spain}

\author{Angelo~Esposito}
\affiliation{Theoretical Particle Physics Laboratory (LPTP), Institute of Physics, EPFL, 1015 Lausanne, Switzerland}

\author{Emma~Geoffray}
\affiliation{Theoretical Particle Physics Laboratory (LPTP), Institute of Physics, EPFL, 1015 Lausanne, Switzerland}

\author{Antonio~D.~Polosa}
\affiliation{Dipartimento di Fisica and INFN, Sapienza Universit\`a di Roma, P.le Aldo Moro 2, I-00185 Roma, Italy}

\author{Sichun~Sun}
\affiliation{Dipartimento di Fisica and INFN, Sapienza Universit\`a di Roma, P.le Aldo Moro 2, I-00185 Roma, Italy}

\begin{abstract}
We consider a model of sub-GeV dark matter whose interaction with the Standard Model is mediated by a new vector boson (the dark photon) which couples kinetically to the photon. We describe the possibility of constraining such a model using a superfluid He-4 detector, by means of an effective theory for the description of the superfluid phonon. We find that such a detector could provide bounds that are competitive with other direct detection experiments only for ultralight vector mediator, in agreement with previous studies. As a byproduct we also present, for the first time, the low-energy effective field theory for the interaction between photons and phonons.
\end{abstract}

\keywords{Light Dark Matter, Effective Theory, Helium, Phonon, Dark Photon}
\pacs{95.35.+d, 67.40.-w, 47.37.+q}

\maketitle

%%%%%%%%%%%%%%%%%%%%%%%%%%%%%%%%%%%%%%%%%%%
%%%%%%%%%%%%%%%%%%%%%%%%%%%%%%%%%%%%%%%%%%%

\section{Introduction}

To understand the origin and nature of dark matter has been a central topic in both theoretical and experimental physics for a long time. In particular, if considered as a new kind of particle, the presence of dark matter would constitute one of the strongest evidences for physics beyond the Standard Model. A large share of the efforts so far has been devoted to the study of the so-called Weakly Interacting Massive Particle, i.e. dark matter particles with masses of order $100$~GeV and interaction strengths comparable to the weak interactions. These searches did not lead to any positive result, yet.

This provides a strong motivation to look into different mass regions, and several efforts have been devoted to the study of sub-GeV dark matter, as proposed by different models --- see e.g.~~\cite{Boehm:2003hm,Boehm:2003ha,Hooper:2008im,Feng:2008ya,Hochberg:2014dra,Kuflik:2015isi,Kaplan:2009ag,Falkowski:2011xh,Hall:2009bx,Chu:2011be,Green:2017ybv,Knapen:2017xzo,Bondarenko:2019vrb,Hong:2019nwd}. To detect such particles experimentally one needs devices with energy thresholds below the eV, and several proposals have been put forth, ranging from superconductors~\cite{Hochberg:2015pha,Hochberg:2016ajh,Hochberg:2019cyy}, polar materials~\cite{Knapen:2017ekk,Griffin:2018bjn,Cox:2019cod}, Dirac materials~\cite{Hochberg:2017wce,Coskuner:2019odd,Geilhufe:2019ndy}, and many others~\cite{Essig:2019xkx,Essig:2019kfe,Trickle:2019ovy,Emken:2019tni,Sanchez-Martinez:2019bac,Bunting:2017net, Essig:2011nj,Essig:2015cda,Essig:2017kqs,Essig:2012yx,Campbell-Deem:2019hdx,Griffin:2019mvc,Trickle:2019nya}.

Among these, the concept of employing a detector based on superfluid He-4 was first presented in~\cite{Lanou:1988,Adams:1996,Guo:2013dt}, and then further developed in~\cite{Schutz:2016tid,Knapen:2016cue,Acanfora:2019con,Caputo:2019cyg}. In particular, the interaction of the dark matter with the bulk of the detector can produce collective excitations, which could then be detected~\cite{Hertel:2018aal,Maris:2017xvi}, allowing a sensitivity to dark matter as light as the keV. If the dark matter interacts with the Standard Model via a scalar mediator, such a detector could provide very promising bounds. In~\cite{Acanfora:2019con,Caputo:2019cyg} the problem has been formulated in terms of a relativistic effective field theory (EFT) for superfluids~\cite{Son:2002zn,Nicolis:2011cs,Nicolis:2015sra}, which allows to describe the interactions of the He-4 phonon with itself and with the dark matter in a simple way, starting from a standard action principle. Such an approach has already been proved to be successful in a number of phenomenological applications --- see e.g.~\cite{Manuel:2011ed,Horn:2015zna,Berezhiani:2015bqa,Esposito:2017xzg,Nicolis:2017eqo,Esposito:2018sdc}.

In this paper we continue this program by studying the case of a sub-GeV dark matter charged under some new $U_\text{d}(1)$ group and interacting with the Standard Model via a new vector mediator (the dark photon) which mixes kinetically with the photon~\cite{Holdom:1985ag,Galison:1983pa}. 

To this end, we write down the most general relativistic low-energy EFT for the interaction between the photon and the bulk of He-4 which, to the best of our knowledge, appears here for the first time. With this at hand, we study the process of emission of a single phonon by the passing dark matter and discuss the result in the context of the present direct, cosmological and astrophysical constraints for the dark photon mass and coupling. In agreement with~\cite{Knapen:2016cue}, we find that a He-4 detector could be competitive with the current bounds for ultra-light dark photons.

\vspace{1em}

\noindent \emph{Conventions:} Throughout this paper we work in natural units, $\hbar=c=\epsilon_0=\mu_0=1$, and adopt a ``mostly plus'' metric signature. Moreover, we use Greek indices to span the full spacetime coordinates and latin indices to span the spatial ones only.

%%%%%%%%%%%%%%%%%%%%%%%%%%%%%%%%%%%%%%%%%%%%
%%%%%%%%%%%%%%%%%%%%%%%%%%%%%%%%%%%%%%%%%%%%

\section{Relativistic EFT for superfluids}

Let us now briefly review the EFT for superfluids, which we will then use to build the most general interaction between the phonon of He-4 and the photon. For an extensive treatment we refer the reader to, for example,~\cite{Nicolis:2011cs,Nicolis:2015sra,Acanfora:2019con}.

From an EFT viewpoint a superfluid is a system characterized by a $U(1)$ internal symmetry associated to a conserved number of particles (e.g. atoms), whose charge $Q$ is at finite density. On top of that, the superfluid \emph{spontaneously} breaks a number of spacetime and internal symmetries, namely boosts, time translations (generated by $H$) and the internal $U(1)$. However, it preserves the combination $\bar H = H-\mu Q$, with $\mu$ being the relativistic chemical potential. Since $H$ is broken the states of the system cannot be classified according to its eigenvalues anymore; one rather needs to use $\bar H$.\footnote{Note, however, that the evolution of the states is still generated by the standard Hamiltonian, $H$~\cite{Nicolis:2011pv,Nicolis:2012vf}.}

The Goldstone boson associated with the above symmetry breaking pattern corresponds to the low-energy collective excitation of the superfluid, i.e. the phonon. The easiest way to implement such a pattern is arguably via a single real scalar field, $\psi(x)$, which shifts under the internal $U(1)$, $\psi\to\psi+a$, and acquires a vacuum expectation value proportional to time, $\langle\psi(x)\rangle = \mu t$. The phonon corresponds to the fluctuation of the field around its equilibrium configuration, $\psi(x)=\mu t+ c_s\sqrt{\mu/\bar n}\,\pi(x)$, where $c_s$ is the superfluid sound speed and $\bar n$ its equilibrium number density. The prefactor has been chosen in order for the field to be canonically normalized.

Given that the breaking is spontaneous, the most general low-energy action for the phonon will have to be invariant under all the above symmetries, and feature the lowest possible number of derivatives. The only possible invariant is $X=\sqrt{-\partial_\mu\psi\partial^\mu\psi}$, which corresponds to the local chemical potential (i.e. in presence of fluctuations). The most general action is~\cite{Son:2002zn,Nicolis:2011cs}
\begin{align} \label{eq:SHe}
\begin{split}
S_\text{eff}&=\int d^4x\,P(X) \\
&=\int d^4x\bigg[\frac{1}{2}\dot\pi^2-\frac{c_s^2}{2}\big(\bm\nabla\pi\big)^2 + \dots\bigg]
\end{split}
\end{align}
Here $P(X)$ is the pressure of the superfluid~\cite{Acanfora:2019con}. For a strongly coupled system like He-4, the analytic form of $P(X)$ is hard to obtain from first principles. Nonetheless, it can be extracted from data~\cite{Abraham:1970aa}, which is the approach adopted here. In the second line of Eq.~\eqref{eq:SHe} we have expanded in small fluctuations around the background. Higher order terms would give all possible self-interactions of the phonon at low energies~\cite{Acanfora:2019con,Caputo:2019cyg}, which will not be necessary for the current study. Indeed, we will focus on the emission of a single phonon, which is the simplest observable and does not involve any further interaction of the phonon with itself.

From Eq.~\eqref{eq:SHe} we see that the dispersion relation for an on-shell phonon is $\omega(q)=c_sq$.
We stress that all the effective couplings are completely fixed by the superfluid equation of state --- e.g. $c_s\equiv c_s(P)$ --- which are extracted directly from data~\cite{Abraham:1970aa}.

The EFT described above is only valid at small momenta, namely when the momenta involved are smaller than a UV cutoff, $\Lambda\sim 1$~keV.\footnote{Alternatively, when the energies are smaller than $c_s\Lambda$.} In particular, this means that it cannot incorporate higher momentum excitations like maxons or rotons. In the rest of this paper we assume to work in this regime. 

Although to have a complete description of all possible excitations one would need to perform a numerical study, we stress that in~\cite{Caputo:2019cyg} it has been shown that the results obtained by means of the EFT match with those obtained with more traditional techniques~\cite{Schutz:2016tid,Knapen:2016cue}. The latter have been tuned on neutron scattering data, and include maxons and rotons as well. It follows that, for the observables of interest, most of the contribution comes from final state phonons, for which the EFT gives an accurate description.

%%%%%%%%%%%%%%%%%%%%%%%%%%%%%%%%%%%%%%%%%%%%
%%%%%%%%%%%%%%%%%%%%%%%%%%%%%%%%%%%%%%%%%%%%

\section{EFT for the interaction between the dark sector and the He-4}

For the sake of clarity we focus on the case of a fermionic dark matter, $\chi(x)$, charged under some dark $U_\text{d}(1)$ group.\footnote{The EFT for the case of a scalar dark matter is only slightly different from the one presented here, and we have checked that it leads to no appreciable differences in the rest of the paper.} As already anticipated, we assume for this particle to interact with the Standard Model via a dark photon, $V_\mu(x)$, which couples to the photon via kinetic mixing, and acquires a mass from some mechanism happening at energy scales much higher than the ones under consideration.

If we assume that the kinetic mixing is the only coupling between the dark sector and the Standard Model this implies that the interaction of the dark matter with He-4 must happen via a dark photon, which then converts into a photon. The low-energy action for the interaction between the photon and the superfluid will have to be invariant under the full Poincar\'e group, under the global $U(1)$ of the superfluid, as well as under the gauge electromagnetic $U_\text{em}(1)$. Moreover, since the He-4 is electrically neutral, it is not possible to build any non-derivative coupling with the photon field, $A_\mu(x)$; the interaction must happen via higher multipoles~\cite{Knapen:2016cue}.

Following these rules, the most general low-energy EFT for the case of interest is described by
\begin{align} \label{eq:Seff}
\begin{split}
S_\text{eff}&=-\int d^4x\bigg[\frac{1}{4}F_{\mu\nu}F^{\mu\nu}+\frac{1}{4}V_{\mu\nu}V^{\mu\nu}-\frac{\epsilon}{2}F_{\mu\nu}V^{\mu\nu} \\
&\quad+\frac{m_V^2}{2}V_\mu V^\mu-\bar \chi\big(i\gamma^\mu D_\mu-m_\chi\big)\chi \\
&\quad-\frac{1}{2}a(X)F_{\mu\nu}F^{\mu\nu}-\frac{1}{2}b(X)F^{\mu\rho}F_{\;\;\rho}^\nu\partial_\mu\psi\partial_\nu\psi\bigg]\,,
\end{split}
\end{align}
where $F_{\mu\nu}$ and $V_{\mu\nu}$ are the field strengths for the photon and dark photon respectively, and the gauge covariant derivative of the dark sector is $D_\mu=\partial_\mu+igV_\mu$. Moreover, we assume for the dark sector to be perturbative, i.e. $g\lesssim4\pi$. Finally, the last line of Eq.~\eqref{eq:Seff} describes the most general coupling between the photon and any number of superfluid phonons at low energies.

The functions $a$ and $b$ are a priori completely generic, i.e. they cannot be found solely on symmetry grounds.\footnote{The procedure presented here corresponds to what, in high energy jargon, is called matching. The UV theory in this case is the description of He-4 in terms of its atomic structure.} However, as we now show, they can be determined in terms of the static properties of the superfluid, namely of its electric and magnetic polarizabilities. Consider the system at equilibrium, $\langle\psi(x)\rangle=\mu t$. In this case the last line of the action~\eqref{eq:Seff} reduces to
\begin{align}
\begin{split}
S_\text{eff}&\supset-\int d^4x\bigg[\frac{1-2a}{4}F_{\mu\nu}F^{\mu\nu}-\frac{\mu^2b}{2}F_{0i}F_{0i}\bigg] \\
&=-\int d^4x\bigg[ \frac{1-2a}{2}\bm B^2-\frac{1-2a+\mu^2b}{2}\bm E^2 \bigg]\,,
\end{split}
\end{align}
where $\bm E$ and $\bm B$ are the electric and magnetic fields, and $a\equiv a(\mu)$ and $b\equiv b(\mu)$ are now evaluated on the background. One recognizes this to be the action for an electromagnetic field in a medium~\cite{Jackson:1998nia}, and therefore the functions $a$ and $b$ can be related to the electric and magnetic polarizabilities, $\alpha_E$ and $\alpha_B$ respectively, by
\begin{align}
1-2a+\mu^2b=1+\alpha_E\bar n\,;\quad \frac{1}{1-2a}=1+\alpha_M\bar n\,.
\end{align}
Since typically $\alpha_M\ll\alpha_E$~\cite{glick1961diamagnetic,fetter1972light}, the effective couplings are given by
\begin{align} \label{eq:ab}
a(\mu)\simeq 0\,;\qquad \mu^2b(\mu)\simeq\bar n(\mu)\alpha_E\,.
\end{align}

The action in Eq.~\eqref{eq:Seff} contains any number of phonons interacting with two photon fields. One can then, in principle, enhance the coupling by introducing an external electric field,\footnote{The introduction of external fields could present experimental difficulties. In this respect, our analysis should be considered as an optimistic one.} $\bar F_{0i}=E_i$, which allows for an interaction term that converts a photon into a phonon,\footnote{One could also introduce a magnetic field. This interaction is however suppressed in the nonrelativistic limit (i.e. for $c_s\ll 1$).} analogous to the Primakoff effect~\cite{Pirmakoff:1951pj}. Indeed, the electric field will induce a polarization of the medium, hence favoring the interaction with the photon. In particular, expanding the last line of Eq.~\eqref{eq:Seff} to linear order in the phonon field, in presence of the external field, one gets
\begin{align}
\begin{split}
S_\text{eff}&\supset\int d^4x\sqrt{\frac{\mu}{\bar n}}c_s\bigg[\bigg(\mu^2\frac{db}{d\mu}+2\mu b\bigg)F_{0i}\dot\pi \\
&\qquad\qquad\qquad\quad+\mu b F_{ij}\nabla_j\pi\bigg]E_i\,.
\end{split}
\end{align}

Everything so far has been general for any electrically neutral s-wave superfluid. He-4 is a nonrelativistic system for which $\mu\simeq m_\text{He}$, $c_s\simeq248$~m/s and $\bar n\simeq8.5\times 10^{22}$~cm$^{-3}$~\cite{Abraham:1970aa}, while the electric polarizability is $\alpha_E\simeq2\times10^{-25}$~cm$^{3}$~\cite{fetter1972light}.

Using Eq.~\eqref{eq:ab}, together with the thermodynamical identities $dP=\bar nd\mu$ and $m_\text{He}c_s^2=dP/d\bar n$, one finds $\mu^2db/d\mu\simeq \frac{\alpha_E\bar n}{m_\text{He}c_s^2}$. Considering that $c_s\ll 1$ and, for an on-shell phonon, $\dot\pi\sim c_s\bm \nabla\pi$,  the photon-phonon interaction in presence of the external electric field can be well approximated by
\begin{align} \label{eq:SApi}
S_\text{eff}\supset\int d^4x \frac{\alpha_E\bar n}{m_\text{He}c_s^2}\sqrt{\frac{m_\text{He}}{\bar n}}c_sE_iF_{0i}\dot\pi\,.
\end{align}
Starting from the actions~\eqref{eq:Seff} and \eqref{eq:SApi} one deduces the following Feynman rules for the dark matter--dark photon interaction, for the dark photon--photon conversion and for the photon--phonon conversion induced by an external $\bm E$-field:
\begin{subequations}
\begin{align}
\includegraphics[width=0.15\textwidth,valign=c]{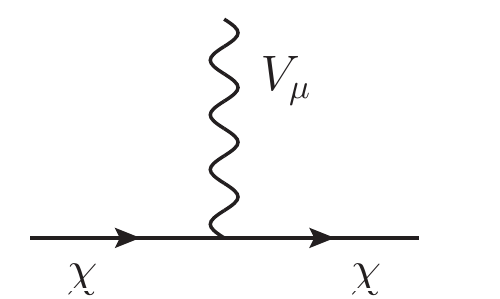}\!\!\!\!\!\!&=-ig\gamma^\mu\,,  \\
\includegraphics[width=0.18\textwidth,valign=c]{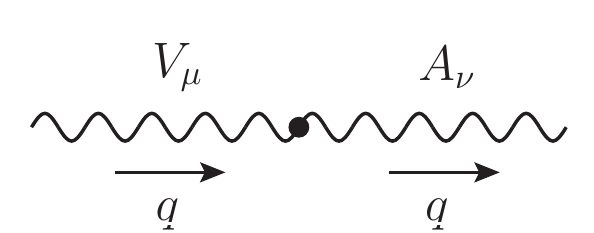}&=i\epsilon\big(q^2\eta^{\mu\nu}-q^\mu q^\nu\big)\,, \\
\includegraphics[width=0.18\textwidth,valign=c]{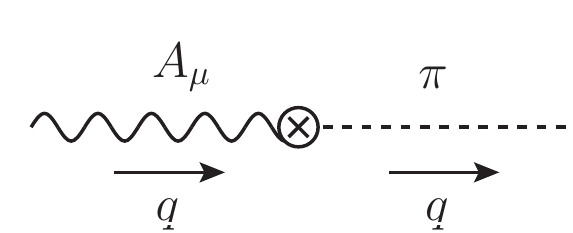}&=i\frac{\bar n\alpha_E}{m_\text{He}c_s}\sqrt{\frac{m_\text{He}}{\bar n}}\times \\ &\quad\times E_i\omega_q\big(\omega_q\delta^\mu_i+q_i\delta^\mu_0\big)\,, \notag
\end{align}
\end{subequations}
where the crossed circle represents the external electric field.
It should be noted that, given the action~\eqref{eq:Seff}, the in-medium photon propagator is modified with respect to the vacuum one. However, the changes are of order $\bar n\alpha_E\sim10^{-2}$. Being a subleading contribution to the matrix element, we will neglect them here.\footnote{For the interested reader, the in-medium photon propagator in Landau gauge reads
\begin{align*}
G_{\mu\nu}(q)&=-\frac{i}{q^2}\bigg[\eta_{\mu\nu}-\frac{q_\mu q_\nu}{q^2} \\
&\quad +\bigg(\delta_\mu^0\delta_\nu^0+\frac{q_0^2}{q^2}\eta_{\mu\nu}+\frac{q_0q_\mu\delta_\nu^0+q_0q_\nu\delta_\mu^0}{q^2}\bigg)\bar n\alpha_E \\
&\quad+O\big(\bar n^2\alpha_E^2\big)\bigg]\,,
\end{align*}
where we have already used Eq.~\eqref{eq:ab}. Note that, since the medium does not break rotations, the inclusion of the above correction in the matrix element does not lead to any new tensor structures and/or anisotropies, as it happens instead for Dirac materials~\cite{Hochberg:2017wce,Coskuner:2019odd,Geilhufe:2019ndy}.}

%%%%%%%%%%%%%%%%%%%%%%%%%%%%%%%%%%%%%%%%%%%%
%%%%%%%%%%%%%%%%%%%%%%%%%%%%%%%%%%%%%%%%%%%%

\section{Phonon emission}

\begin{figure}[t!]
\centering
\includegraphics[width=0.2\textwidth]{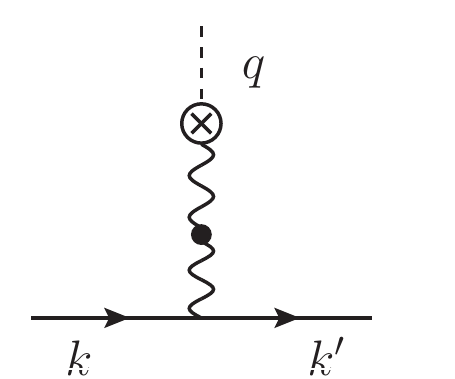}
\caption{Feynman diagram for the amplitude of emission of a single phonon in presence of an external electric field that polarizes the medium.} \label{fig:diagram}
\end{figure}

In this work we focus on the simplest possible process, namely the emission of a single phonon after the interaction of the dark matter with the bulk of He-4. The amplitude of interest is given by the Feynman diagram in Figure~\ref{fig:diagram}. Averaging over the initial dark matter polarizations and summing over its final ones, one gets
\begin{align}
\frac{1}{2}\sum_\text{spin}\big|\mathcal{M}\big|^2\simeq\frac{4m_\chi^2\bar n\alpha^2_E\epsilon^2g^2}{m_\text{He}}\bigg(\frac{|\bm q|\bm q\cdot\bm E}{\bm q^2+m_V^2}\bigg)^2\,,
\end{align}
where we have used the nonrelativistic limit for the dark matter, $k^{(\prime)}\simeq(m_\chi,\bm 0)$, and for the He-4, $c_s\ll1$.
The corresponding rate is
\begin{align} \label{eq:dgamma}
\frac{d\Gamma}{d\omega}&=\frac{\bar n \alpha_E^2\epsilon^2g^2}{4\pi m_\text{He}v_\chi c_s^2}\bm{E}^2\bigg(\frac{\omega^2}{\omega^2+c_s^2m_V^2}\bigg)^2\cos^2\theta_E\,,
\end{align}
where the angle between the incoming dark matter and the outgoing phonon is fixed by kinematics to be on the Cherenkov cone, $\cos\theta=\frac{c_s}{v_\chi}+\frac{q}{2m_\chi v_\chi}$, with $v_\chi$ the dark matter velocity~\cite{Acanfora:2019con,Caputo:2019cyg}. 
Moreover, $\cos\theta_E=\cos\theta\cos\theta_\chi-\cos(\phi-\phi_\chi)\sin\theta\sin\theta_\chi$ is the angle between the electric field and the outgoing phonon, with $(\theta_\chi,\phi_\chi)$ the angle between the incoming dark matter and the electric field.

The rate of events per unit target mass is obtained as
\begin{align} \label{eq:R}
R=\int dv_\chi\,f_\text{MB}(v_\chi)\frac{\rho_\chi}{m_\text{He}\bar n m_\chi}\int_{\omega_\text{min}}^{\omega_\text{max}}d\omega \frac{d\Gamma}{d\omega}\,.
\end{align}
The maximum energy that the dark matter can transfer to a phonon is either fixed by kinematics (namely requiring $\cos\theta<1$) or by the cutoff of the EFT, and it is $\omega_\text{max}=\min\big(2m_\chi c_s(v_\chi-c_s),c_s\Lambda\big)$.\footnote{Note that consistency with the regime of applicability of the EFT does not limit the dark matter mass, but rather only the exchanged momentum. A heavy dark matter can still softly scatter off of the He-4 detector so to excite a phonon degree of freedom.} On the other hand, the outgoing phonon must have energy larger than a certain value in order for it to be detected. When only a single phonon is involved, it cannot release to the system enough energy to induce an appreciable change in temperature in the detector~\cite{Hertel:2018aal}. It can, however, be observed via the so-called ``quantum evaporation'', which sets the minimum energy to be the binding energy of a He-4 atom to the rest of the bulk, $\omega_\text{min}=0.62$~meV~\cite{Maris:2017xvi}. In particular, given the value of $\omega_\text{max}$, this implies that the final state phonon will be detectable only for a dark matter heavier than roughly $0.1$~MeV.

\begin{figure}[t]
\centering
\includegraphics[width=0.45\textwidth]{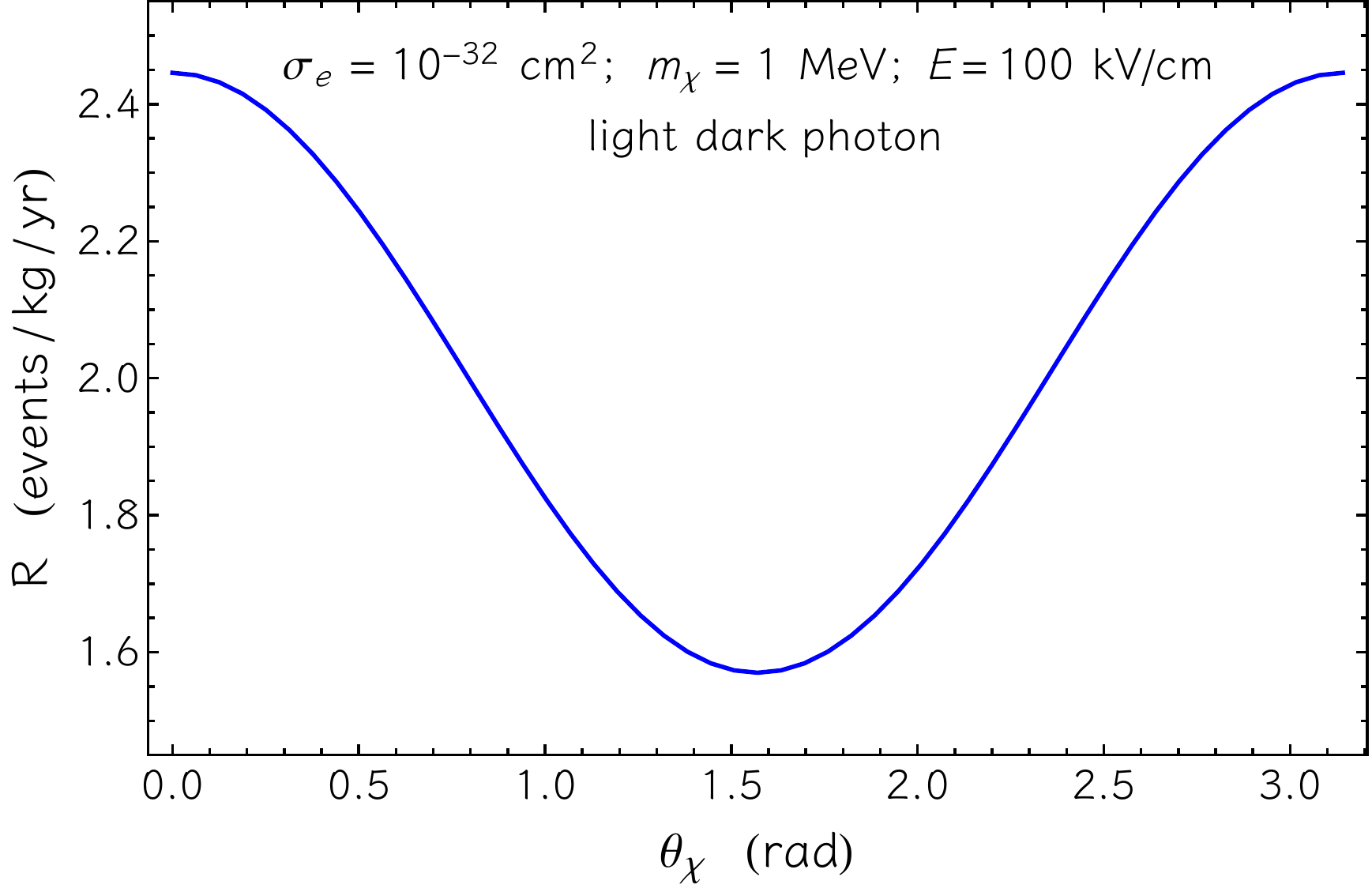}
\caption{Rate of events per unit time and detector mass as a function of the relative polar angle between the incoming dark matter and the external electric field. The plot has been obtained in the approximation of light dark photon, $m_V\ll q$.} \label{fig:modulation}
\end{figure}

Importantly, the rate in Eq.~\eqref{eq:dgamma} depends on the relative angle between the direction of the incoming dark matter and the electric field, as can also be seen in Figure~\ref{fig:modulation}. This induces a sensible modulation in the number of events. By suitably rotating the external field with time, one could employ this to discriminate signal from background~\cite{Knapen:2016cue}.

%%%%%%%%%%%%%%%%%%%%%%%%%%%%%%%%%%%%%%%%%%%%
%%%%%%%%%%%%%%%%%%%%%%%%%%%%%%%%%%%%%%%%%%%%

\section{Projections and comparison with existing bounds}

\begin{figure}[t!]
\centering
\includegraphics[width=0.48\textwidth]{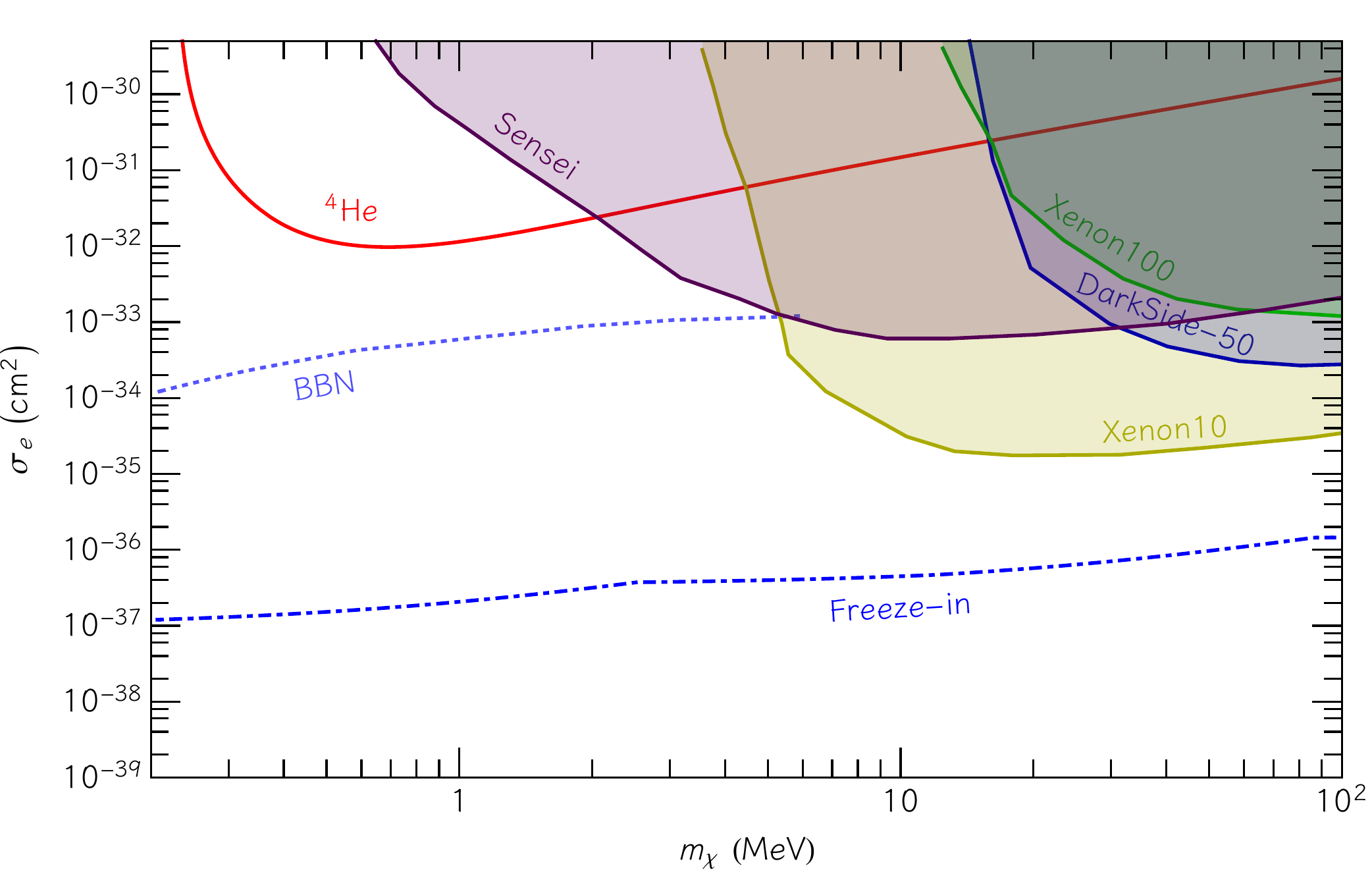}
\caption{Projected constraints on the electron cross section as a function of the dark matter mass for a He-4 experiment, as compared to the existing bounds from SENSEI~\cite{Abramoff:2019dfb}, XENON10, XENON100~\cite{Essig:2017kqs} and DarkSide-50~\cite{Agnes:2018oej}. We consider a 95\% C.L. for a year of exposure and a kg of material, assuming zero background. We also report the current BBN bounds~\cite{Davidson:2000hf}, as well as a combination of mass and cross section that would allow to explain the dark matter relic abundance via a freeze-in mechanism~\cite{McDonald:2001vt,Hall:2009bx,Bernal:2017kxu}.} \label{boundcross}
\end{figure}

Starting from Eq.~\eqref{eq:R} we can compute the projected excluded region. In particular, we will consider an external electric field $E=100$~kV/cm, which has been shown to be realistically achievable in lab~\cite{Ito:2015hwa}.\footnote{We have checked that for the case of the emission of two phonons, in the configuration where the $\bm E$-field is parallel to the exchanged momentum, we recover the bounds presented in~\cite{Knapen:2016cue}.} 

There are two distinct scenarios here: the heavy dark photon case ($m_V\gg |\bm q|$) and the light dark photon one ($m_V\ll |\bm q|$). In the former we find that the best sensitivity one can achieve is 
\begin{align}
g\epsilon\simeq10^{-9}\bigg( \frac{m_V}{1\text{ keV}} \bigg)^2 \qquad\text{for} \qquad m_\chi\simeq1\text{ MeV}\,,
\end{align}
which is already largely excluded by the existing stellar and accelerator constraints --- see e.g.~\cite{Lin:2019uvt}.

For ultra-light dark photon, instead, most of the cosmological and astrophysical bounds can be evaded when $m_V\lesssim10^{-14}$~eV.
In this case, we can use Eq.~\eqref{eq:R} to compute the expected sensitivity for the dark matter-electron cross section, which can be written as 
\begin{align}
	\sigma_e\equiv \frac{4 g^2 \epsilon^2}{\alpha_\text{em}^3m_e^4}\frac{m_e^2 m_\chi^2}{(m_e+m_\chi)^2}\,,
\end{align}
where $\alpha_\text{em}$ is the fine-structure constant and $m_e$ the electron mass. 
In Figure~\ref{boundcross} we show our results as compared to other direct detection experiments~\cite{Abramoff:2019dfb,Essig:2017kqs}. As one can see, a He-4 detector could be competitive in the sub-MeV region. Note that the masses excluded by He-4 would fall in the region already excluded by Big Bang Nucleosynthesis (BBN) constrains~\cite{Davidson:2000hf, Vogel:2013raa}. The same region would also be covered by the SN1987A supernova bound~\cite{Davidson:2000hf,Chang:2016ntp} on which, however, some doubts have been recently casted~\cite{Bar:2019ifz}. 

We also show the curve for dark matter relic abundance via freeze-in scenario~\cite{McDonald:2001vt,Hall:2009bx,Bernal:2017kxu,Dvorkin:2019zdi}, i.e. a scenario in which  the interaction is very weak and slowly builds up the dark matter relic abundance non-thermally.  In the case of light mediator, $m_{V}\ll m_{\chi}$, the dark matter production cross section from fermions $f$ of the Standard Model thermal bath, $f\bar{f} \rightarrow\chi \bar{\chi}$, reads
\begin{align}
	\langle\sigma v\rangle \propto \frac{\alpha_{\chi}\alpha_{f}}{T^2}\,.
\end{align}
where $\alpha_{\chi}\equiv \frac{g^2}{4\pi}, \alpha_{f}\equiv \frac{\epsilon^2e^2}{4\pi}$. 
The relic abundance is then 
\begin{align}
	\Omega_{\chi} h^2 \simeq 0.12 \bigg(\frac{\alpha_{\chi}}{10^{-13}}\bigg)\bigg(\frac{\alpha_{f}}{10^{-14}}\bigg)\bigg(\frac{1 \text{ MeV}}{m_{\chi}}\bigg)^2\,.
\end{align}
For each value of the mass $m_{\chi}$ the value of the dark matter relic abundance uniquely fixes the combination $\alpha_{\chi}\alpha_{f} \propto \epsilon^2 g^2$.

%%%%%%%%%%%%%%%%%%%%%%%%%%%%%%%%%%%%%%%%%%%%
%%%%%%%%%%%%%%%%%%%%%%%%%%%%%%%%%%%%%%%%%%%%

\section{Conclusion}

In this work we have studied the response of a He-4 detector to the interaction of a sub-GeV dark matter particle which interacts with the Standard Model via a dark photon, kinetically mixed with the photon. In order to do that, we have employed a relativistic EFT to describe the low-energy interactions of the superfluid phonon with the dark matter. On top of that, we also presented the most general coupling between the photon and the bulk of He-4.

We considered the simplest possible process, i.e. the emission of a single phonon by the passing dark matter, whose rate can be enhanced by introducing an external electric field. For a dark matter lighter than the MeV such an observable could be competitive with the existing direct detection experiments, although that region should already be excluded by BBN bounds.

The case of a two-phonon final state has already been discussed in~\cite{Knapen:2016cue}. As already commented, we have checked that using our EFT we recover the same results.

In conclusion, in addition to the strong bounds that a He-4 detector could put on the parameters of a dark matter interacting with the Standard Model via a scalar mediator, it can also provide some important information in the sub-MeV region for a dark matter mediated by a dark photon.

Moreover, as already shown in~\cite{Acanfora:2019con,Caputo:2019cyg}, the superfluid EFT approach proves to be particularly clean and clear to tackle a particle physics problem like the one at hand.

%%%%%%%%%%%%%%%%%%%%%%%%%%%%%%%%%%%%%%%%%%%%
%%%%%%%%%%%%%%%%%%%%%%%%%%%%%%%%%%%%%%%%%%%%

\begin{acknowledgments}
We are grateful to S.~Baracchini, K.~Blum, G.~Cavoto, S.~de~Cecco, S.~Knapen, T.~Lin, H.~Liu, T.~Melia, R.~Rattazzi, D.~Redigolo, J.~Redondo and T.~Volansky for useful discussions and comments on the draft. A.E. and E.G. are supported by the Swiss National Science Foundation under contract 200020-169696 and through the National Center of Competence in Research SwissMAP. A.C. is supported by ``Generalitat Valenciana'' (Spain) through the ``plan GenT'' program (CIDEGENT/2018/019), by grants FPA2014-57816-P, PROMETEOII/2014/050 and SEV-2014-0398, as well as by the EU projects H2020-MSCA-RISE-2015 and H2020-MSCA-ITN- 2015//674896-ELUSIVES.  S.S. was supported by MIUR in Italy under Contract (No. PRIN 2015P5SBHT) and ERC Ideas Advanced Grant (No. 267985) \textquotedblleft DaMeSyFla". 
\end{acknowledgments}

\bibliographystyle{apsrev4-1}
\bibliography{biblio}

\end{document}